\documentclass[runningheads]{llncs}
\usepackage{multirow}
\usepackage{graphicx}
\begin{document}
\title{DeepFEL: Deep Fastfood Ensemble Learning for Histopathology Image Analysis}
\titlerunning{DeepFEL: Deep Fastfood Ensemble Learning for Histopathology}
%


\author{Nima Hatami}
\authorrunning{N. Hatami}

\institute{CREATIS, CNRS UMR5220, INSERM U1206, Université Lyon 1, INSA-Lyon, France \\
\email{hatami@creatis.insa-lyon.fr}}


\maketitle              

\begin{abstract}

Computational pathology tasks have some unique characterises such as multi-gigapixel images, tedious and frequently uncertain annotations, and unavailability of large number of cases \cite{madabhushi2016image}. To address some of these issues, we present {\em Deep Fastfood Ensembles} -- a simple, fast and yet effective method for combining deep features pooled from popular CNN models pre-trained on totally different source domains (e.g., natural image objects) and projected onto diverse dimensions using random projections, the so-called Fastfood \cite{Fastfood}. The final ensemble output is obtained by a consensus of simple individual classifiers, each of which is trained on a different collection of random basis vectors. This offers extremely fast and yet effective solution, especially when training times and domain labels are of the essence. We demonstrate the effectiveness of the proposed deep fastfood ensemble learning as compared to the state-of-the-art methods for three different tasks in histopathology image analysis. 

\end{abstract}
\section{Introduction}
Growing digitalization of histopathology slides has led to computational pathology playing an increasingly important role by using the power of deep learning, image analytics and big data to enhance diagnostic precision \cite{niazi2019digital,bera2019artificial}. However, there are several challenges that have arisen in the relatively new field of computational pathology. Histology images are gigantic compared to standard computer vision images. A single Whole Slide Image (WSI) of a tissue can be as big as 100,000$\times$100,000 pixels with 10GB of memory storage requirements. Therefore, data annotation and training of standard convolutional neural networks (CNNs) can take days and sometimes weeks. Another challenge is lack of enough samples from specific classes. Sometimes, there are not so many samples available from rare abnormalities and diseases, compared to the normal cases. Another challenge is that there may not be so many annotations available by the expert pathologists \cite{madabhushi2016image}. Furthermore, many publicly available datasets (such as TCGA \cite{TCGA}) have only a small proportion of diagnostic slides available for a particular organ. Due to all these reasons, performance of the standard data-hungry CNNs drops caused by small sample sizes (despite using transfer learning, since the target domain is so different from the source domain).

Random Kitchen Sinks or RKS \cite{Rahimi1,Rahimi2} is a kernel based method that approximates the function $f$ by means of multiplying the input with a Gaussian random matrix, followed by the application of a nonlinearity. Fastfood approximation \cite{Fastfood} accelerates RKS $100\times$ faster and using $1,000\times$ less memory. This is achieved by replacing Gaussian matrices in RKS by Hadamard matrices and scaling matrices that are inexpensive to multiply and store. However, these methods belong to the ``pre-deep learning era", and models such as deep CNNs have introduced the new state-of-the art results in many computer vision tasks by leveraging on the availability of large volumes of data and more powerful hardware, but often requiring long training times.  

In this paper, we introduce Deep Fastfood Ensemble Learning (DeepFEL) as a fast, simple and yet effective approach for computational pathology. It applies histopathology images (target domain) and pools multiple sets of deep features from the popular CNN models trained on totally different source domain (e.g. natural images) and projected onto multiple diverse dimensions. Sets of independent simple classifiers each trained on a particular set of random vectors are generated. The final ensemble result is obtained by consensus of the individual classifiers. Combining the power of deep features offered by pre-trained CNNs and computational efficiency offered by Fastfood approximation, we achieve fast, yet effective classifier ensemble, especially for cases with limited number of samples. 

The rest of the paper is organized as follows: the deep Fastfood ensemble learning is presented in the next section. Datasets description and experimental results are in section \ref{Experiments}. Finally, section \ref{Conclusions} concludes the paper. 

\section{Deep Fastfood Ensemble Learning}

\begin{figure}[!t]
\centering
\includegraphics[scale=0.45]{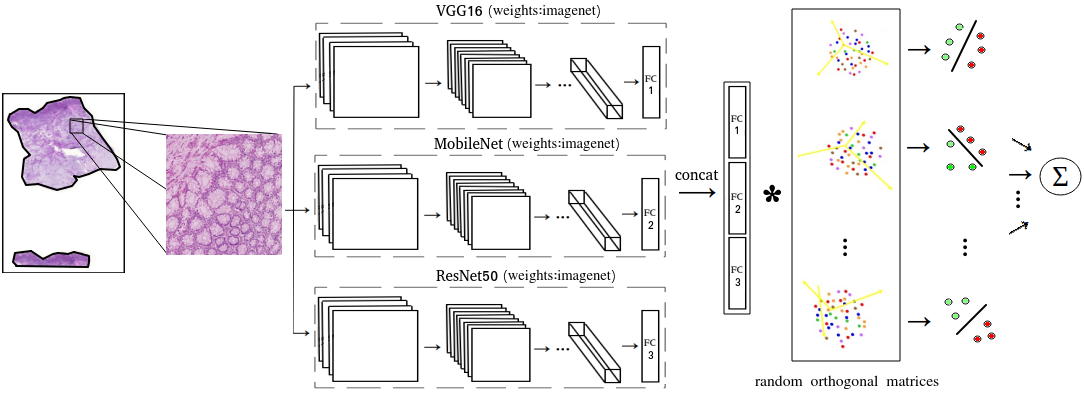}
\caption{Block diagram of the proposed deep FEL for fusion of the transferred deep features. The backbone models (e.g. VGG16, MobineNet, and ResNet50) are only trained on ImageNet (source domain), and have not seen the target domain i.e. histopathology images.} 
\label{fig_fastfood_ens}
\end{figure}

A block diagram of the proposed Deep Fastfood Ensemble Learning (DeepFEL) method is shown in Figure \ref{fig_fastfood_ens}. First a WSI is pre-processed, the tissue is segmented, and the image patches are extracted from the tissue area. Each patch is fed into multiple CNNs of diverse architecture and weights. The deep features are then pooled from the last layers of the networks. The deep features are then concatenated into a single vector of dimension $1\times M$. In order to have an ensemble of $N$ classifiers each working on $D$-dimensional feature space (where $D<<M$), $N$ matrices of $M\times D$ are generated by Fastfood approximation so that the random matrices are mutually orthogonal. After projection of an image into $N$ $D$-dimensional spaces, $N$ simple classifier is being trained, each one specialized on a separate space. The final ensemble output is obtained by a consensus of the individual classifiers, e.g. weighed average of the outputs of the individual classifiers. 

The proposed DeepFEL offers three main advantages: ({\em i}) it uses features obtained from popular deep CNN models pre-trained on completely different problems (source domain) and combining them to solve problems in computational pathology (target domain), thus absolving the need to spend days of training time; ({\em ii}) it is time and memory efficient and yet obtains reasonable performance using ordinary hardware; and ({\em iii}) unlike deep learning models that are data-hungry, it performs satisfactorily with a relatively small number of samples. This is especially critical in medical image analysis domain, where a few samples of rare abnormalities are available or there is a need for specialized pathologists to do a large number of annotations. 

As far as Random Sink Kitchens and its Fastfood approximation are concerned, the proposed DeepFEL is a continuation of these methods with the following changes: instead of raw image or its hand-crafted features, here we use combination of transferred deep features pooled from multiple CNN architectures. From a deep learning perspective, DeepFEL makes use of existing CNN models pre-trained on different source domains (e.g. natural images) and provides an efficient approach to fuse them in order to solve computational pathology tasks. 

There are two main ingredients of a successful ensemble: (a) good accuracy of individual member classifiers, so that each could solve an aspect of the task as accurately as possible and (b) diversity among the members, so that they each cover different parts of the problem. In DeepFEL, we achieve high accuracy due to the quality of deep features, and quality of the projection into lower-dimensional spaces. The diversity, however, comes from architecture and weight differences between CNNs that the deep features are pooled from, and also randomness of the projection matrices and the fact that they are mutually orthogonal. 

\section{Experimental Results and Discussion}
\label{Experiments}

In this section, we first describe the datasets that are used to carry out the experiments, and then report the experimental settings and present the results. 

\subsection{Datasets}
Three different histopathology image datasets consisting of digitized images of H\&E-stained glass slides of cancerous tissue specimens with different characteristics are used to investigate the performance of the proposed deep Fastfood ensembles. Table \ref{table_datasets} briefly lists the main characteristics of each dataset. 

\begin{table}[!t]
\renewcommand{\arraystretch}{1.3}
\caption{Datasets used in the paper. The table lists characteristics of the datasets and the total numbers of training, validation, and test images.}
\label{table_datasets}
\centering
\begin{tabular}{|c|c|c|c|}
\hline
Dataset & CRC-TIA & PCam & NCT-CRC-HE \\
\hline
\# of patches & 36,750 & 327,680 & 107,180 \\
patch dimension & 1750 px$^{2}$ & 96 px$^{2}$ & 224 px$^{2}$  \\
\# of slides  & 139 & 400 & 136 \\
\# of classes & 4 & 2 & 9  \\
magnification & 20x & 10x & 20x \\
organ & colorectal & lymph node & colorectal \\
problem & cancer grading & non/tumor & tissue types\\
\hline
\end{tabular}
\end{table}

\begin{figure}[!t]
\centering
\includegraphics[scale=0.4]{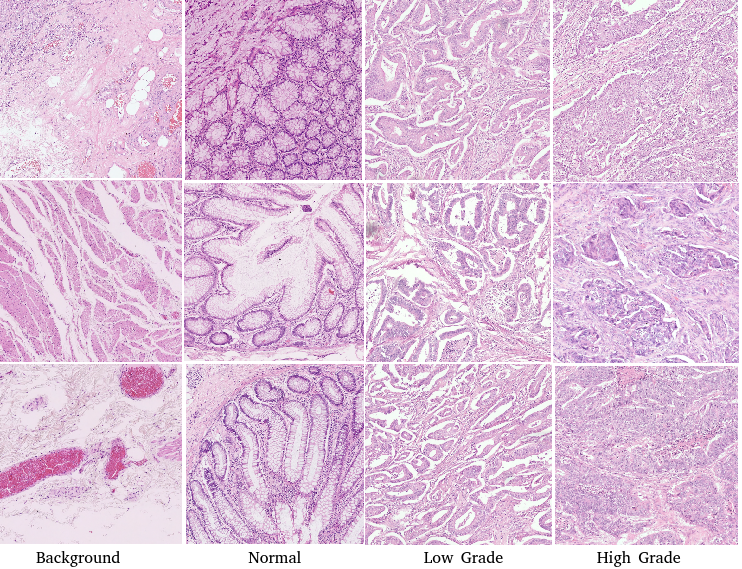}
\caption{Samples from 4 classes of the Colorectal Cancer Grading (CRC-TIA) Dataset.} 
\label{fig_crc-tia}
\end{figure}

\textbf{CRC-TIA}\cite{CRC_TIA1,CRC_TIA2}:
The dataset comprises of 36,750 non-overlapping images of size 1,750$\times$1,750 pixels, extracted at magnification 20$\times$. Each image is labelled as normal, low grade tumor or high grade tumor by an expert pathologist. To obtain these images, digitized WSIs of 139 CRC tissue slides stained with H\&E are used. All WSIs were taken from different patients and were scanned using the Omnyx VL120 scanner at 0.275 $\mu$m/pixel (40$\times$). Samples of the dataset are shown in Figure \ref{fig_crc-tia}.

\begin{figure}[!t]
\centering
\includegraphics[scale=0.2]{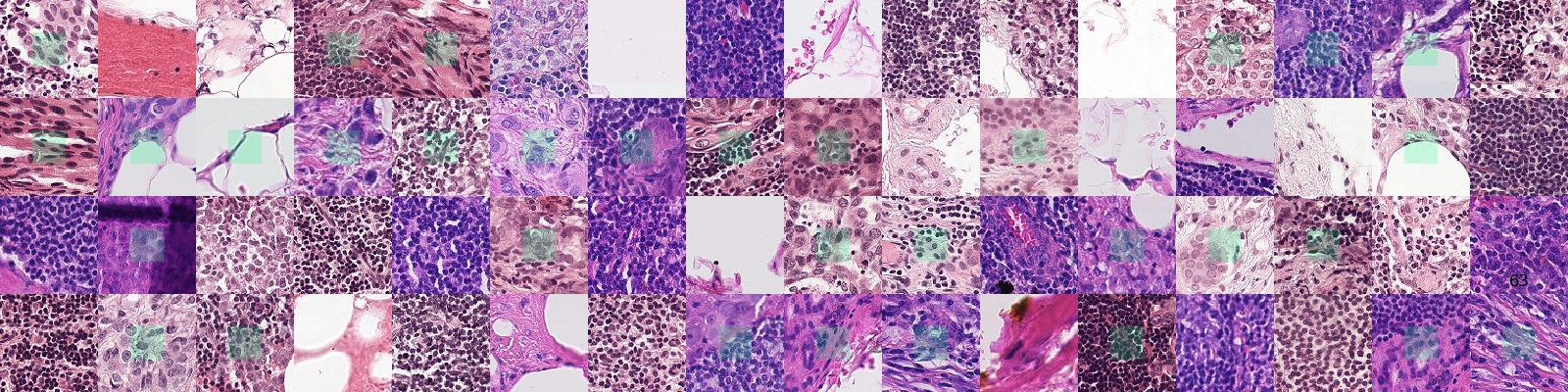}
\caption{Examples from PCam data. Green boxes indicate tumor images.} 
\label{fig_PCam}
\end{figure}

\textbf{PCam}\cite{PCam}: The PatchCamelyon metastasis detection benchmark consists of 327,680 color images (96$\times$96px) extracted from digital scans of breast cancer lymph node sections. 
PCam is derived from the Camelyon16 Challenge \cite{camelyon16}, which contains 400 H\&E stained WSIs of sentinel lymph node sections. The slides were acquired and digitized at 2 different centers using a 40$\times$ objective. We then undersample images at 10$\times$ to increase the field of view. We follow the train/test split from the Camelyon16 challenge, and further hold-out 20\% of the train WSIs for the validation set. To prevent selecting background patches, slides are converted to HSV, blurred, and patches filtered out if maximum pixel saturation lies below 0.07. Each image is given a binary label indicating presence of metastatic tissue. Samples of the dataset are shown in Figure \ref{fig_PCam}.

\begin{figure}[!t]
\centering
\includegraphics[scale=0.5]{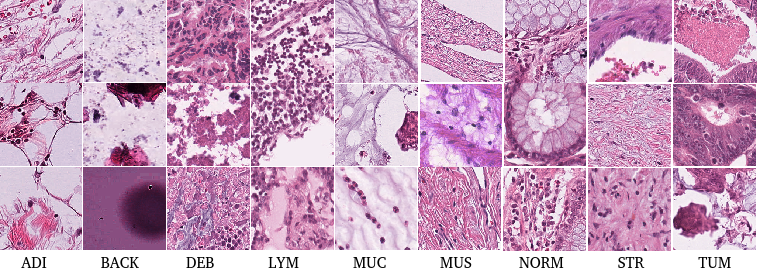}
\caption{Samples from the NCT-CRC-HE-100K dataset; the 9 tissue classes are: Adipose (ADI), background (BACK), debris (DEB), lymphocytes (LYM), mucus (MUC), smooth muscle (MUS), normal colon mucosa (NORM), cancer-associated stroma (STR) and colorectal adenocarcinoma epithelium (TUM).} 
\label{fig_kather100k}
\end{figure}

\textbf{NCT-CRC-HE}\cite{kather100_data}:
The training data (NCT-CRC-HE-100K) is a set of 100,000 non-overlapping image patches from H\&E-stained histological images of human colorectal cancer (CRC) and normal tissue. All images are 224$\times$224 px at 0.5 microns per pixel (MPP). All images are color-normalized using Macenko's method \cite{macenko2009}. Tissue classes are: Adipose (ADI), background (BACK), debris (DEB), lymphocytes (LYM), mucus (MUC), smooth muscle (MUS), normal colon mucosa (NORM), cancer-associated stroma (STR) and colorectal adenocarcinoma epithelium (TUM). The images were manually extracted from 86 H\&E stained cancer tissue slides from formalin-fixed paraffin-embedded (FFPE) samples from the NCT Biobank (National Center for Tumor Diseases, Heidelberg, Germany) and the UMM pathology archive (University Medical Center Mannheim, Mannheim, Germany). Tissue samples contained CRC primary tumor slides and tumor tissue from CRC liver metastases; normal tissue classes were augmented with non-tumorous regions from gastrectomy specimen to increase variability. The validation data (NCT-CRC-HE-7K) is a set of 7,180 image patches from N=50 patients with colorectal adenocarcinoma (no overlap with patients in NCT-CRC-HE-100K). Samples of the dataset are shown in Figure \ref{fig_kather100k}.

\subsection{Settings and Results}
We carried out experiments in TensorFlow.Keras using Python. Six popular CNN models VGG16\cite{VGG16}, VGG19\cite{VGG16}, Xception\cite{Xception}, ResNet50\cite{ResNet}, MobileNet\cite{MobileNet}, and DenseNet\cite{DenseNet} with pre-trained ImageNet weights are used as baselines and also as backbones for extracting deep features for the DeepFEL. In order to obtain the best baseline results, SGD and Adam optimizers with learning rates of 0.01 and 0.001 are tried with number of epochs 20 and batch size of 64. Horizontal and vertical flips and rotation range of 40 degrees are applied for data augmentation. For DeepFEL, features from ``global\_average\_pooling2d'' layer are pooled from each model. DeepFEL has two parameters to set: the dimension of projections ($D$) and the number of ensemble members (N). In our experiments, we explored $D:\{20,100,1K, 10K, 20K\}$ and $N:\{10,20,50\}$, and the best $\{D,N\}$ is selected based on cross-validation performance. Simple Decision Trees \cite{DecisionTree} classifiers are used for the ensemble. 

\begin{table}[!t]
\renewcommand{\arraystretch}{1.3}
\caption{Performance (in terms of accuracy \% and time in minutes) of DeepFEL as compared to different popular models on CRC-TIA cancer grading dataset with different training sizes ($n$ samples per class).}
\label{table_CRC_TIA}
\centering
\begin{tabular}{|c|c|c|c|c|c|c|}
\hline
Model & \multicolumn{2}{c}{Number of samples per class}   &\\
 & n=50 & n=100 & full data (n=3,500) \\
\hline
VGG16 & 70.06\% (1.5min)  & 81.85\% (3min) & 90.08\% (15min) \\
VGG19 & 70.55\% (1.5min)  & 81.33\% (3min) & 89.85\% (16min) \\
Xception & 71.98\% (1.5min)  & 82.09\% (3.1min) & \textbf{91.19}\% (16min)\\
ResNet50 & 70.25\% (1.5min)  & 83.10\% (3min) & 90.64\% (17min)\\
MobileNet & 73.56\% (1min)  & 82.60\% (2.8min) & 90.33\% (17min) \\
DenseNet & 75.88\% (1.7min)  & 82.05\% (3min) & 90.50\% (16min) \\
\bf{Ours} & \bf{77.80\% (0.5min)}  & \bf{83.65\% (1.2min)} & 90.85\% \textbf{(8min)}\\
\hline
\end{tabular}
\end{table}

Tables \ref{table_CRC_TIA}, \ref{table_PCam} and \ref{table_kather100} report the experimental results. Different training sizes of $n:50$, $n:100$ and full data ($n$: number of images per class) is used to evaluate the performance of the models, and measure their sensitivity towards data size. Performance is measured in terms of accuracy \% and time (total training and test) in minutes. As shown, all three tables are following the same pattern. When smaller number of training samples is available ($n:50,100$), the Fastfood ensemble outperforms other models both in terms of accuracy and time. When there are enough samples for training, the end-to-end CNNs are winner (with a marginal difference) in terms of accuracy. However, Fastfood ensemble is still a good choice when the time is an essence.

\begin{table}[!t]
\renewcommand{\arraystretch}{1.3}
\caption{Performance (in terms of accuracy \% and time in minutes) of DeepFEL, compared to different popular models on PCam data with different training sizes ($n$: samples per class).}
\label{table_PCam}
\centering
\begin{tabular}{|c|c|c|c|c|c|c|}
\hline
Model & \multicolumn{2}{c}{Number of samples per class}   &\\
 & n=50 & n=100 & full data (n$\sim$160K) \\
\hline
VGG16 & 67.55\% (6.5min)  & 78.11\% (12min) & 88.48\% (62min) \\
VGG19 & 67.07\% (6.5min)  & 77.98\% (13min) & 88.55\% (66min) \\
Xception & 67.08\% (6.5min)  & 79.05\% (14.1min) & 88.79\% (66min)\\
ResNet50 & 68.55\% (6.5min)  & 77.65\% (13min) & 89.67\% (70min)\\
MobileNet & 67.98\% (6min)  & 77.60\% (12min) & \textbf{89.83\%} (67min) \\
DenseNet & 69.18\% (7min)  & 76.82\% (13min) & 89.18\% (66min) \\
\bf{Ours} & \bf{71.84\% (2min)}  & \bf{78.95\% (5min)} & 88.13\% \textbf{(32min)}\\
\hline
\end{tabular}
\end{table}

\begin{table}[!t]
\renewcommand{\arraystretch}{1.3}
\caption{Performance (in terms of accuracy \% and time in minutes) of DeepFEL, as compared to different popular models on NCT-CRC-HE tissue types data with different training sizes ($n$ samples per class).}
\label{table_kather100}
\centering
\begin{tabular}{|c|c|c|c|c|c|c|}
\hline
Model & \multicolumn{3}{c|}{Number of samples per class}   \\ 
 & n:50 & n:100 & full data (n$\sim$11K) \\
\hline
VGG16 & 79.86\% (2min)  & 82.85\% (7min) & 94.48\% (45min) \\
VGG19 & 78.49\% (2min)  & 81.34\% (7min) & 94.35\% (46min) \\
Xception & 64.87\% (2min)  & 81.99\% (8min) & 93.19\% (47min)\\
ResNet50 & 35.25\% (2min)  & 60.20\% (7min) & 93.62\% (48min)\\
MobileNet & 68.16\% (2min)  & 82.70\% (6min) & 94.40\% (49min) \\
DenseNet & 79.82\% (2.5min)  & 83.95\% (7min) & \textbf{95.10\%} (46min) \\
\bf{Ours} & \bf{80.02\% (0.6min)}  & \bf{84.70\% (3.7min)} & 93.10\% \textbf{(22min)}\\
\hline
\end{tabular}
\end{table}

\section{Conclusions}
\label{Conclusions}

In this paper, we presented Deep Fastfood Ensemble Learning (DeepFEL) as a method for fusion of deep features projected onto lower dimensions for computational pathology. This simple and fast, yet effective approach takes advantage of recent progresses in deep learning and kernel methods into one framework. The experiments are carried out on three computational pathology tasks, i.e. CRC grading, tumor metastasis detection, and tissue type classification. The results suggest two main conclusions: i) when small number of training samples are available, DeepFEL is more accurate and faster than its CNN rivals; and ii) when large number of samples available, DeepFEL obtains similar accuracy as compared to the end-to-end deep CNNs, but at a much reduced computational cost.  



\bibliographystyle{splncs04}
\bibliography{refs.bib}
\end{document}